\documentclass[aps, prb, reprint, superscriptaddress]{revtex4-2}

\usepackage{graphicx}
\usepackage{bm}
\usepackage{bbm}
\usepackage{amsmath}

\usepackage[usenames,dvipsnames]{color}
\usepackage[bbgreekl]{mathbbol}
\usepackage{xcolor}

\newcommand{\calT}{\mathcal{T}}
\newcommand{\calB}{\mathcal{B}}

\begin{document}
\title{Two-wire junction of inequivalent Tomonaga-Luttinger liquids}

\author{Yao-Tai Kang}
\affiliation{ School of Electronic and Electrical Engineering, Zhaoqing University, Zhaoqing 526061, China}
\affiliation{ Department of Physics, National Tsing Hua University, Hsinchu 30013, Taiwan}

\author{Chung-Yu Lo}
\affiliation{ Department of Physics, National Tsing Hua University, Hsinchu 30013, Taiwan}

\author{Masaki Oshikawa}
\affiliation{Institute for Solid State Physics, University of Tokyo, Kashiwa 277-8581, Japan}

\author{Ying-Jer Kao }
\email{yjkao@phys.ntu.edu.tw}
\affiliation{ Department of Physics, National Taiwan University, Taipei 10607, Taiwan}
\affiliation{ Physics Division, National Center for Theoretical Sciences, Taipei 10617, Taiwan}

\author{Pochung Chen}
\email{pcchen@phys.nthu.edu.tw}
\affiliation{ Department of Physics, National Tsing Hua University, Hsinchu 30013, Taiwan}
\affiliation{ Physics Division, National Center for Theoretical Sciences, Taipei 10617, Taiwan}
\affiliation{ Frontier Center for Theory and Computation, National Tsing Hua University, Hsinchu 30013, Taiwan}

\date{\today}

\begin{abstract}
  We develop two novel numerical schemes to study the conductance of the two-wire junction of inequivalent Tomonaga-Luttinger liquids.
  In the first scheme we use the static current-current correlation function across the junction to extract the linear conductance through a relation that is derived via the bosonization method.
  In the second scheme we apply a voltage bias and evaluate the time-dependent current across the junction to obtain the current-voltage characteristic.
  The conductance is then extracted from the small bias result within the linear response regime.
  Both schemes are based on the infinite-size matrix product state to minimize the finite-size effects.
  Due to the lack of the translational invariance, we focus on a finite-size window containing the junction.
  For time-independent calculations, we use infinite boundary conditions to evaluate the correlations within the window.
  For time-dependent calculations, we use the window technique to evaluate the local currents within the window.
  The numerical results obtained by both schemes show excellent agreement with the analytical predictions.
\end{abstract}

\maketitle

\graphicspath{{figures/}}

\section{introduction}

Transport properties of the strongly correlated quasi-one-dimensional (1D) quantum systems 
have been the subject of intensive investigation in recent years due to the potential applications in nanoelectronics.
In these systems, electron-electron interaction has drastic effects and the Fermi liquid theory breaks down.
Instead, the system is described by the Tomonaga-Luttinger liquid (TLL) theory,
which is parameterized by a Luttinger parameter $g$ \cite{Tomonaga:1950eu, Luttinger:1963iw, Giamarchi:2004uc}.
Experimentally, TLL's characteristic behavior has been observed experimentally in a variety of quasi-1D systems 
\cite{Bockrath.1999, Yao:1999jn,Postma:2000jy, Ishii:2003ipa,Kim:2007ixa}.
In this work we focus on an important class of the quasi-1D transport problem: junctions of multiple TLL wires.
The simplest setup is a two-wire junction of equivalent TLL.
It consists of two TLL wires with the same Luttinger parameter connected by a weak link.
Theoretically, it is well known that in this case, the system renormalizes either 
to the single fully connected wire fixed point or to the two disconnected wires fixed point, 
depending on the sign of the interaction \cite{Kane:1992kx, Kane:1992hd, Furusaki:1993fk,Rylands:2016eu}.
For three-wire Y junctions of equivalent TLL, more conductance fixed points begin to emerge \cite{Egger:20036b8, Chamon:2003cp, Oshikawa:2006il}.
From the perspective of the experiment, it is also important to study the influence of the contact as well as the Fermi liquid leads \cite{Janzen.2006} and the multiwire junction with inequivalent TLLs \cite{Safi:1995bra, Hou:2012hz, Lal:2002kt, BarnabeTheriault:2005eaa, Hou:2012hz}.

In the above-mentioned studies, the bosonization method has been used extensively to draw important conclusions \cite{Kane:1992kx, Kane:1992hd, Furusaki:1993fk,  Furusaki:1993sb, Safi:1995bra, Chamon:2003cp, Oshikawa:2006il, Hou:2012hz, Sedlmayr:2012eh, Sedlmayr:2014bia}.
However, in order to go beyond the perturbative regime to reach a comprehensive understanding of quasi-1D systems' transport properties,
other methods are called for. 
Analytically, an exact solution method based on Bethe ansatz has been developed and applied successfully to several systems
\cite{Mehta.2006,Rylands:2016eu, Rylands:2017cr}.
However, it is restricted to integrable models. 
On the other hand,  many numerical methods have been developed.
Within fermion representation, methods based on renormalization group equations \cite{Aristov:2009jo,  Aristov:2010gu, Aristov:2011fj} and functional renormalization group (fRG) technique \cite{Andergassen:2004cz, Meden:2008bi} have been developed to evaluate the one-particle Green's function from which the linear conductance can be extracted.
Numerical renormalization group (NRG)  method \cite{Wilson.1975}, 
which is originally developed for the equilibrium properties of quantum impurity systems,
has also been generalized to study the transport properties of nanodevices with noninteracting leads 
\cite{Cornaglia.2004, Anders.2008, Anders.2008wfu}.
However, it is difficult to generalize the NRG based method to study a junction with interacting leads.

The method developed here is based on the density matrix renormalization group (DMRG) technique,
which is a powerful and versatile numerical tool to study quasi-1D systems \cite{Schollwock:2011gl}.
Over the years, several DMRG based methods have been developed to study the transport properties of quasi-1D systems.
One of the earliest approaches uses a ring geometry and extracts the conductance from the current induced by the flux
\cite{Molina.2004, Meden.2003z5a, Meden.2003}.
Another approach uses the linear response theory to relate the conductance and the dynamical correlation functions,
which can be evaluated via the correction vector DMRG method \cite{Bohr:2006hz} 
or the dynamical DMRG method \cite{Bohr:2006hz, Bischoff:2019kq, Branschadel2010}.
In Refs.~\cite{Rahmani:2010jr, Rahmani:2012bq, Hou:2012hz} a general method to extract the conductance tensor of the multiwire junction is proposed and is used to study the multiwire junction of equivalent  and inequivalent TLLs.
By using boundary conformal field theory, the conductance tensor is related to the static correlation function of a semi-infinite system.
A conformal transformation is then used to connect the correlators of a finite system to a  semi-infinite one,
making it possible to extract the conductance tensor from the static current-current correlator of a finite system.
We note in passing that in this approach it is necessary to add a mirror-image junction during the transformation.
However, the Hamiltonian of the mirror-image junction can be rigorously derived only for the case of noninteracting wires.
On the other hand, since the invention of time-dependent DMRG (tDMRG), 
various approaches have been developed to simulate the time-dependent current of the multiwire junction,
from which the linear response and the full current-voltage characteristics can be obtained
\cite{Schmitteckert:2004jx, Schneider:2006tn, Boulat:2008hx, AlHassanieh:2006fd, Shunsuke:2008jr, daSilva:2008fw, Feiguin:2008jd, HeidrichMeisner:2009gv, HeidrichMeisner:2009jg,HeidrichMeisner:2010fr}.
Recently, a relation between the static charge correlations and the linear has also been put forward \cite{He.2021}.
Typically, a finite-size system with open boundary condition is simulated. However, this may lead to strong finite-size effects.
In order to reduce the finite-size effects, modified boundary conditions have also been explored 
\cite{Schneider:2006tn, AlHassanieh:2006fd, Bohr:2007, Feiguin:2008jd}.

In this work, we develop two novel numerical schemes to study the linear and the nonlinear conductance of the multiwire junctions.
Our main strategy is to always work with an infinite system to minimize the finite-size effects but only perform measurements within a finite size window to make the simulation feasible.
This also removes the need to perform the conformal transformation to obtain an effective finite size system
and the addition of the mirror-image junction.
Specifically, we revisit the problem of the two-wire junction of inequivalent TLLs to benchmark our methods.
It is known that in this case the conductance is determined by a single effective Luttinger parameter
\cite{Safi:1995bra, Hou:2012hz, Lal:2002kt, BarnabeTheriault:2005eaa, Hou:2012hz}.
In the first scheme, we use the method recently proposed by some of us in Ref.~\cite{Lo:2019ki} 
to calculate the static current-current correlation function of the two-wire junction
and use the method proposed in Refs.~\cite{Rahmani:2010jr, Rahmani:2012bq, Hou:2012hz} to extract the linear conductance.
However, we find that the key relation between the conductance and the static current-current correlation function
derived in Refs.~\cite{Rahmani:2010jr, Rahmani:2012bq, Hou:2012hz} needs minor modification when the Luttinger parameters on two wires are different.
While we only measure the correlation function within a finite size window, 
we show that a moderate window size already allows us to observe the asymptotic behavior.
Our results agree excellently with the theoretical prediction and verify that the conductance is indeed governed by the effective Luttinger parameter.

To probe the nonlinear conductance, we incorporate the window technique developed in Ref.~\cite{Zauner:2015jl}
to evaluate directly the local currents within a finite size window, after a source-drain bias is applied to the system. 
We show that after the transient time, we can obtain a very flat quasistationary current 
up to a time scale that is limited by the window size and the carrier velocity.
This allows us to define an average current with very small error,
from which the current-voltage characteristics can be obtained. 
Our results show a wide range of linear response regime,
and the linear conductance can be extracted from a small bias calculation.
Furthermore, we verify that the linear conductance obtained from the nonequilibrium setup 
is highly consistent with the results via static correlations.

The rest of the paper is organized as follows.
In Sec.~\ref{sec:model}, we set up the notation and define the Hamiltonian of the model.
In Sec.~\ref{sec:bosonization}, we use the bosonization method to derive the modified key relationship 
between the conductance and the static current-current correlation function.
In Sec.~\ref{sec:method}, we discuss the main ingredients of our numerical method and outline the steps.
In Sec.\ref{sec:static}, we present our numerical results from the time-independent calculation,
while in Sec.\ref{sec:time} we present our numerical results from the time-dependent calculations.
Finally we summarize in Sec.\ref{sec:sum} and discuss future directions.


\section{Model}
\label{sec:model}

\begin{figure}[t]
  \includegraphics[width=\columnwidth]{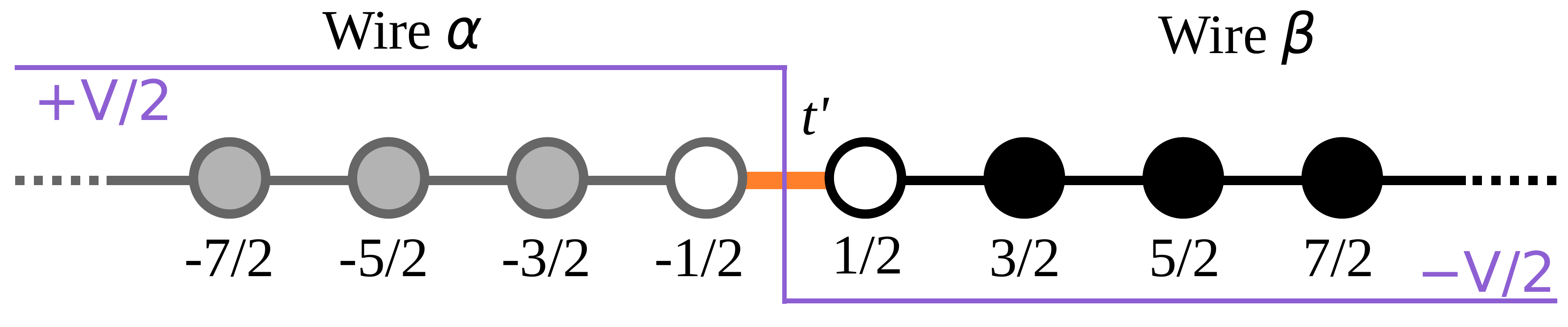}
  \caption{Sketch of the two-wire junction and the bias profile.}
  \label{fig:model}
\end{figure}

We consider a two-wire junction which consists of two semi-infinite long TLL wires connected by a weak link as sketched in Fig.~\ref{fig:model}.
To model such a junction,  we start from two semi-infinite long wires with the Hamiltonian
\begin{equation}
  H^{\mu}_{\text{wire}} = \sum_{\substack{i \in \mathcal{Z}{\color{red}^{\ge}} + \frac{1}{2} }}
  -\left( c^{\mu \dagger}_{i} c^\mu_{i+1} + H.c. \right) + U^{\mu} \tilde{n}^\mu_{i} \tilde{n}^\mu_{i+1},
\end{equation}
where $c^{\mu \dagger}_i (c^\mu_i)$ with $\mu \in \alpha, \beta$ is the creation (annihilation) operator at the site $i$ of the wire $\mu$
and $\tilde{n}^\mu_i \equiv c^{\mu \dagger}_{i} c^\mu_{i} - \frac{1}{2}$.
$\mathcal{Z}^{\ge}$ denotes the set of non-negative integers.
We note that the hopping strength is set to unity as the energy scale.
Furthermore, $H^{\mu}_{\text{wire}}$ can also be expressed in the form of a translational invariant matrix product operator (MPO) \cite{Pirvu:2010ky}
\begin{equation}
  H^{\mu}_{\text{wire}} = \cdots W^{\mu}_{-\frac{3}{2}} W^{\mu}_{-\frac{1}{2}} W^{\mu}_{\frac{1}{2}} W^{\mu}_{\frac{3}{2}} \cdots,
\end{equation}
where 
\begin{equation}
  \label{MPO_bulk}
  W^{\mu}_i =  W^{\mu} = 
  \left[
    \begin{array}{ccccc}
    \mathbbm{1}_i & 0 & 0 & 0 & 0 \\
    c^\mu_i & 0 & 0 & 0 & 0 \\
    c^{\mu \dagger}_i & 0 & 0 & 0 & 0 \\
    \tilde{n}^\mu_i & 0 & 0 & 0 & 0  \\
    0 & -c^{\mu}_i & -c^{\mu  \dagger}_i & U\tilde{n}^\mu_i & \mathbbm{1}_i \\
    \end{array}
  \right].
\end{equation}
When $ |U^{\mu} | < 2$, the wire is in the TLL phase.
At half filling, its Luttinger parameter $g^\mu$ is determined by $U^{\mu}$ through the relation
\begin{equation}
  g^{\mu} =\frac{\pi}{2\arccos(-U^{\mu}/2)},
\end{equation}
while the carrier velocity reads 
\begin{equation}
  v^{\mu} =\pi  \frac{\sqrt{1-(U^{\mu} /2)^2}}{\arccos(U^{\mu} /2)}.
\end{equation}
We note that $U^\mu=0$ corresponds to a noninteracting wire with $g=1$,
while positive and negative interaction correspond to $g>1$ and $g<1$, respectively.

We form a two-wires junction by connecting 
a semi-infinite long wire with label $\alpha$ extending to the left and
a semi-infinite long wire with label $\beta$ extending to the right
at site $i=\pm 1/2$ by a link of strength $t^\prime$.
The Hamiltonian of the link reads:
\begin{equation}
  H_{\text{link}}  = -t^\prime \left( c^{\alpha \dagger}_{-1/2} c^\beta_{1/2} + c^{\beta \dagger}_{1/2} c^\alpha_{-1/2} \right),
\end{equation}
while the Hamiltonian of the junction reads
\begin{eqnarray}
    \label{eq:H_junc}
    & & H_{\text{junc}} = H_{\text{link}} \nonumber \\
    &+& \sum_{\substack{i \in \mathcal{Z}^{\ge} + \frac{1}{2} }}   
    -\left( c^{\alpha \dagger}_{-i} c^\alpha_{-(i+1)} + H.c. \right) + U^\alpha \tilde{n}^\alpha_{-i} \tilde{n}^\alpha_{-(i+1)} \nonumber \\
    &+& \sum_{\substack{i \in \mathcal{Z}^{\ge} + \frac{1}{2} }}   
    -\left( c^{\beta \dagger}_{i} c^\beta_{i+1} + H.c. \right) + U^\beta \tilde{n}^\beta_{i} \tilde{n}^\beta_{i+1},
\end{eqnarray}
where $\mathcal{Z}^{\ge}$ denotes the set of non-negative integers.
In the form of the MPO the $H_{\text{junc}}$ is expressed as
\begin{equation}
  H_{\text{junc}} = \cdots W^{\alpha}_{-\frac{5}{2}} W^{\alpha}_{-\frac{3}{2}} \widetilde{W}_{-\frac{1}{2}} W^{\beta}_{\frac{1}{2}} W^{\beta}_{\frac{3}{2}} \cdots,
\end{equation}
where $\widetilde{W}_{-\frac{1}{2}}$ is associated with the $H_{\text{link}}$ as follows:
\begin{equation}
  \label{MPO_junc}
  \widetilde{W}_{-1/2} = 
  \left[
    \begin{array}{ccccc}
    \mathbbm{1}_{-1/2} & 0 & 0 & 0 & 0 \\
    c^\alpha_{-1/2} & 0 & 0 & 0 & 0 \\
    c^{\alpha \dagger}_{-1/2} & 0 & 0 & 0 & 0 \\
    \tilde{n}^\alpha_{-1/2} & 0 & 0 & 0 & 0  \\
    0 & -t^\prime c^{\alpha \dagger}_{-1/2} & -t^\prime  c^\alpha_{-1/2} & 0 & \mathbbm{1}_{-1/2} \\
    \end{array}
  \right].
\end{equation}
In the following we denote the ground state of $H_{\text{junc}}$ by $|\Psi_{\text{junc}}\rangle$.
We define the current operator $J(r)$ as
\begin{equation}
  J(r) =
  \begin{cases}
    i\left( c^{\alpha \dagger}_{r-\frac{1}{2}} c^\alpha_{r+\frac{1}{2}} - c^{\alpha \dagger}_{r+\frac{1}{2}} c^\alpha_{r-\frac{1}{2}} \right)=J^\alpha(-|r|), \; r<0 \\
    i\left( c^{\beta \dagger}_{r-\frac{1}{2}} c^\beta_{r+\frac{1}{2}} - c^{\beta \dagger}_{r+\frac{1}{2}} c^\beta_{r-\frac{1}{2}} \right)=J^\beta(+|r|), \; r > 0 \\
    i t^\prime \left( c^{\alpha \dagger}_{-\frac{1}{2}} c^\beta_{\frac{1}{2}} - c^{\beta \dagger}_{\frac{1}{2}} c^\alpha_{-\frac{1}{2}} \right)=J_{\text{link}} , \; r=0,
   \end{cases}
\end{equation}
where we define the current operator across the link as $J_{\text{link}}$.
Furthermore, the static current-current correlation function is defined as
\begin{equation}
  \langle J(-r) J(r) \rangle_{\text{junc}} \equiv \langle \Psi_{\text{junc}} | J(-r) J(r) |\Psi_{\text{junc}} \rangle.
\end{equation}
It is also convenient to define a $r$-dependent conductance 
\begin{equation}
  \label{eq:G_static_r}
 G^{\alpha \beta}(r) \equiv - \frac{e^2 (2 \pi^2)}{h}
  \left(\frac{v^\alpha + v^\beta}{v^\alpha v^\beta} \right)^2 r^2 \langle J(-r) J(r) \rangle_{\text{junc}}.
\end{equation}
It will be shown in Sec.~\ref{sec:bosonization} that $G^{\alpha\beta}(r)$ approaches
the linear conductance $G^{\alpha\beta}$ as $r$ goes to infinity, i.e.,
\begin{equation}
  \label{eq:G_static}
 G^{\alpha \beta} = \lim_{r \rightarrow \infty}  G^{\alpha \beta}(r).
\end{equation}
We note that Eq.~(\ref{eq:G_static_r}) is different from the result in Refs.~\cite{Rahmani:2012bq, Hou:2012hz} except when $v^\alpha=v^\beta$.
In Sec.~\ref{sec:static} we shall report numerical results that support Eqs.~(\ref{eq:G_static_r}) and (\ref{eq:G_static}).

To probe the nonlinear response and determine the current-voltage characteristics,
we turn on a bias at time $t=0$ and apply a voltage of $\pm V/2$ to the left and right wires, respectively.
We then evaluate the time-dependent local current
\begin{equation}
  \langle J(r, t) \rangle_V = \frac{2\pi e}{h} \langle \Psi_{\text{junc}}(V, t)| J(r) | \Psi_{\text{junc}}(V, t)\rangle,
\end{equation}
where
\begin{equation}
   | \Psi_{\text{junc}}(V, t)\rangle = e^{i(H_{\text{junc}}+H_{\text{bias}}(V))t} |\Psi_{\text{junc}}\rangle
\end{equation}
and
\begin{equation}
  H_{\text{bias}}(V) = \frac{V}{2}  \sum_{\substack{i \in \mathcal{Z}^{\ge} + \frac{1}{2} }}  \left( \tilde{n}^\alpha_{-i} - \tilde{n}^\beta_{i} \right)
\end{equation}
is the bias Hamiltonian.
Formally we define the steady state current across the link as
\begin{equation}
  \label{eq:J}
  J(V) = \lim_{t\rightarrow \infty} \langle J_{\text{link}}(t) \rangle_V
  =\lim_{t\rightarrow \infty} \langle J(r=0, t) \rangle_V.
\end{equation}
Numerically it is difficult to reach the infinite time limit.
In practice we expect that after a transient time the local current across the link will become quasistationary,
from which we can define a time-averaged local current
\begin{equation}
  \label{eq:J_junc}
  \bar{J}_{\text{link}}(V) = \frac{1}{t_2-t_1} \int_{t_1}^{t_2} dt \langle J_{\text{link}}(t) \rangle_V
\end{equation}
within a time window $[t_1, t_2]$. Here we show explicitly the $V$ dependence for clarity.
If one can simulate accurately the local current up to a large enough time to contain a window 
such that that $t_1$ is larger than the transient time scale and $t_2-t_1$ is large enough to obtain a good average,
then the current-voltage characteristics can be accurately extracted.
Furthermore, the linear conductance $G^{\alpha \beta}$ can be obtained by
\begin{equation}
  G^{\alpha \beta} = \lim_{V \rightarrow 0} \frac{\bar{J}_{\text{link}}(V)}{V}
\end{equation}
from a time-dependent calculation. 
We note in passing that this provides a consistency check by comparing with results from the time-independent calculations.

\section{Bosonization}
\label{sec:bosonization}

In this section we first review known results in the literature,
then we derive Eqs.~(\ref{eq:G_static_r}) and (\ref{eq:G_static}) via the bosonization method.
It is shown in Ref.~\cite{Safi:1995bra} that the conductance $G^{\alpha \beta}$ of a junction of inequivalent TLLs
depends only on an effective Luttinger parameter $g_\text{e}$, where
\begin{equation}
  \frac{1}{g_{\text{e}}} = \frac{1}{2}\left( \frac{1}{g^\alpha}+\frac{1}{g^\beta} \right).
\end{equation}
Combined with the result for a junction of equivalent TLLs \cite{Kane:1992kx, Kane:1992hd} one finds:
When $g_{\text{e}}>1$ the conductance takes a universal value 
\begin{equation}
	G^{\text{th}}=g_{\text{e}} \frac{e^2}{h}, \; g_{\text{e}}>1
  \label{eq:G_D}
\end{equation}
regardless the link strength of the junction. In contrast, when $g_{\text{e}}<1$ one has
\begin{equation}
	G^{\text{th}}=0, \; g_{\text{e}}<1.
  \label{eq:G_N}
\end{equation}
It is worth commenting on the case of $g_{\text{e}}=1$.
In general, the effective theory is given by the boundary sine-Gordon theory with the marginal boundary interaction.
For the free fermion ($g^\mu=1$), the coefficient of the boundary interaction can be determined exactly 
and it is related to the exact conductance. This results in \cite{Kane:1992kx, Kane:1992hd}
\begin{equation}
  \label{eq:G_g1}
  G^{\text{th}}=\frac{e^2}{h} \frac{4(t^\prime)^2}{(1+(t^\prime)^2)^2}, \; g^\mu=1.
\end{equation}
In the presence of the interaction ($g^\mu \neq 1$), the effective theory is still the same if $g_{\text{e}}$ is still 1.
While there is no reason to expect a strong renormalization of the coefficient of the marginal boundary interaction, 
the conductance should stay approximately the same as in the free fermion limit.
The conductance is, however, eventually determined by the nonuniversal coefficient of the operator in the field theory.
We hence suspect that the coefficient is renormalized weakly by the interaction, resulting in a weak change of the conductance.
It will been shown later that our numerical results do support such a scenario.
Unfortunately, there is no simple way to determine analytically the weak change of the coefficient and the conductance using the field theory.

In the literature, the relation between the conductance and the static current-current correlation function has been
derived within the framework of boundary conformal field theory for the case of
(i) multiple wires with the same Luttinger parameter and carrier velocity \cite{Rahmani:2010jr},
(ii) multiple wires with the same Luttinger parameter but different carrier velocities \cite{Rahmani:2012bq}, and
(iii) multiple wires with different Luttinger parameters and different carrier velocities \cite{Hou:2012hz}.
We note that the form of the velocity dependence in Ref.~\cite{Rahmani:2012bq} and Ref.~\cite{Hou:2012hz} is the same,
and it falls back to the results in Ref.~\cite{Rahmani:2010jr} if all the velocities are the same.
However, as will be shown below, we find that this velocity dependence needs to be modified.
On the other hand, in Ref.~\cite{Hou:2012hz} it is shown that by rescaling the bosonic fields,
the junction of two wires with different Luttinger parameters
can be mapped to a junction with a single effective Luttinger parameter $g_e$, in agreement with Ref.~\cite{Safi:1995bra}.
We confirm that this stands valid even when the two wires have different charge velocities, because the charge velocities
can be absorbed by a proper rescaling.

We first follow the derivation in Refs.~\cite{Rahmani:2012bq, Hou:2012hz} to obtain the result for the case of $v^\alpha=v^\beta=1$.
We then pay special attention to the case of inequivalent TLLs with different Fermi velocities
and identify the proper way to rescale the equation.
We start from the Kubo formula for the conductance Eq.~(\ref{eq.JJtau0})
\begin{multline}
  G^{\alpha \beta} = \lim_{\omega \to 0_+} 
  -\frac{e^2}{h} \frac{1}{\omega l}
  \int_{-\infty}^\infty d\tau \; e^{i \omega \tau} \\
  \times \int_{l_1}^{l_2} dx \;
  \langle \calT J^\alpha(y,\tau) J^\beta(x,0) \rangle,
\label{eq.Kubo}
\end{multline}
where the electric field is applied uniformly on the finite segment $l_1 < -y, x < l_2$ of the infinite system,
and $l=l_2-l_1$ is the length of the segment.
Here $\calT$ indicates imaginary-time ordering.
When the two wires are inequivalent with two different Luttinger parameters $g_\alpha \neq g_\beta$ but still $v_\alpha = v_\beta=1$,
\begin{multline}
 \langle \calT J^\alpha(z_1,\bar{z}_1) J^\beta(z_2,\bar{z}_2) \rangle \\
  =
\frac{g_{\mathrm{e}}}{4\pi^2}
\left[
A^{\alpha \beta}_\calB
\frac{1}{(\bar{z}_1 - z_2)^2}
+
A^{\beta \alpha}_\calB
\frac{1}{(z_1 - \bar{z}_2)^2}
\right] ,
\label{eq.JJ-corr}
\end{multline}
where
$z \equiv \tau + ix, \bar{z} \equiv \tau -ix$, and
the coefficients $A^{\alpha \beta}_\calB$ are universal and
determined by the conformally invariant boundary condition on the real axis.
By using the identity
\begin{equation}
  \int_{-\infty}^\infty d\tau \;
  \frac{e^{i\omega \tau}}{(\tau - i u)^2}
  = - 2\pi \omega H(u) e^{-\omega \tau},
\end{equation}
where $H(u)$ is the Heaviside step function, 
and Eq.~(\ref{eq.JJ-corr}) (with $z_1=\tau-iy$ and $z_2=ix$),
Eq.~\eqref{eq.Kubo} is reduced to
\begin{align}
 G^{\alpha \beta} & =
\frac{g_{\mathrm{e}} e^2}{h}
\frac{1}{l} \int_{l_1}^{l_2} dx \;
\left[
A^{\alpha \beta}_\calB
H(x-y)
+
A^{\beta \alpha}_\calB
H(-x+y)
\right] 
\\
& =
\frac{g_{\mathrm{e}} e^2}{h}
A^{\alpha \beta}_\calB .
\end{align}
This relates the conductance with the universal coefficient $A^{\alpha \beta}_\calB$ for each
conformally invariant boundary condition.

For a nonchiral (time-reversal invariant) junction,
\begin{equation} 
 A^{\alpha \beta}_\calB = A^{\beta \alpha}_\calB ,
\end{equation}
and thus the conductance is related to the asymptotic behavior
of the current-current correlation function as
\begin{equation}
 G^{\alpha \beta} = G^{\beta \alpha} \sim - \frac{e^2}{h}(8 \pi^2)
 r^2 \langle J^\alpha(-r) J^\beta(r) \rangle ,
\end{equation}
for sufficiently large $r$.
In fact, in the present problem of the junction of two wires, for generic values of the Luttinger parameters,
we only need to consider the Dirichlet boundary condition 
\begin{equation}
  A^{\alpha \beta}_\calB = A^{\beta \alpha}_\calB = 1
\label{eq:A_Dirichlet}
\end{equation}
and the Neumann boundary condition 
\begin{equation}
  A^{\alpha \beta}_\calB = A^{\beta \alpha}_\calB = 0 .
\end{equation}
They are stable when $g_{\text{e}} > 1$ and $g_{\text{e}} < 1$, resulting in the conductance as in Eqs.~\eqref{eq:G_D} and~\eqref{eq:G_N}, respectively.
Because the asymptotic current-current correlation function is dominated by the subleading corrections in the latter case, 
in this paper we are mostly interested in the Dirichlet boundary condition~\eqref{eq:A_Dirichlet}
which represent maximally conducting junction realized for $g_{\text{e}} > 1$.
So far we have just followed Refs.~\cite{Rahmani:2012bq, Hou:2012hz}.

Now let us resurrect the carrier velocity of each wire.
When the velocity is $v$,
the holomorphic/antiholomorphic variables $z,\bar{z}$ should
be proportional to $v \tau \pm i x$.
In other words, we can either
\begin{itemize}
 \item define the rescaled coordinate $\tilde{x} = x/v$, so that
$z = \tau + i \tilde{x}$, or
 \item define the rescaled time $\tilde{\tau} = v \tau$, so that
$z = \tilde{\tau} + i x$.
\end{itemize}
For a single, uniform wire, these two formulations are equivalent under a rescaling
by the factor of $v$.
However, when multiple inequivalent wires are coupled, they
are not equivalent.
In our problem, the wires are coupled at a single junction, and both wires share the same time. 
Namely, the electron hops from one of the wires at time $t$ should appear on the other side of the junction at the same time $t$.
If one scales the time differently for the two wires, the electron transfer at the junction becomes nonlocal in the rescaled time variable.
We should better avoid such a complication and instead rescale the spatial coordinate of each wire to
define holomorphic/antiholomorphic variables:
\begin{align}
 z_1 & \equiv \tau_1 + i \frac{-y}{v^\alpha}, \; \bar{z}_1 \equiv \tau_1 - i \frac{-y}{v^\alpha} \\
 z_2 & \equiv \tau_2 + i \frac{x}{v^\beta}, \; \bar{z}_2 \equiv \tau_2 - i \frac{x}{v^\beta}.
\end{align}
This means that the same holomorphic/antiholomorphic coordinate corresponds to different distances from 
the origin (junction). However, this does not pose a problem, as the two wires are connected only at the junction.

With this rescaling, the current $J$ remains unchanged,
while the charge density $\rho$ in each wire is
multiplied by the velocity $v^\mu$.
This is because the current measures the charge passing
through a specific point per unit time, while the charge density
measures the charge per unit length.
Only the latter is renormalized by the rescaling of the length.

Thus, including the velocity, the current-current correlation
function Eq.~\eqref{eq.JJ-corr} appearing in the Kubo formula Eq.~\eqref{eq.Kubo}
reads (with $\tau_1=\tau$ and $\tau_2=0$)
\begin{widetext}
\begin{align}
\langle \calT J^\alpha(y, \tau) J^\beta(x,0) \rangle 
=
\frac{g_{\mathrm{e}}}{4\pi^2}
\left[
A^{\alpha \beta}_\calB
\frac{1}{\left( \tau - i (\frac{-y}{v^\alpha} + \frac{x}{v^\beta})\right)^2}
+
A^{\beta \alpha}_\calB
\frac{1}{\left( \tau + i (\frac{-y}{v^\alpha} + \frac{x}{v^\beta})\right)^2}
\right] .
\label{eq.JJ-xytau}
\end{align}
Upon integration, one has
\begin{align}
 G^{\alpha \beta} & =
\frac{g_{\mathrm{e}} e^2}{h}
\frac{1}{l} \int_{l_1}^{l_2} dx \;
\left[
A^{\alpha \beta}_\calB
H(\frac{x}{v^\beta}+\frac{-y}{v^\alpha})
+
A^{\beta \alpha}_\calB
H(-\frac{x}{v^\beta}-\frac{-y}{v^\alpha})
\right] 
\\
& =
\frac{g_{\mathrm{e}} e^2}{h}
A^{\alpha \beta}_\calB .
\end{align}
\end{widetext}

Thus the conductance remains the same universal value independent of the velocities $v^\alpha, v^\beta$.
On the other hand, the real-space correlation function depends on the velocity factor.
The equal-time correlation function of currents is given
by setting $\tau=0$ in Eq.~(\ref{eq.JJ-xytau}) as
\begin{align}
\label{eq.JJtau0}
 \langle  J^\alpha(y, 0) J^\beta(x,0) \rangle
& \sim 
- \frac{g_{\mathrm{e}}}{4\pi^2}
\left( A^{\alpha \beta}_\calB + A^{\beta \alpha}_\calB \right)
\frac{1}{ \left( \frac{-y}{v^\alpha} + \frac{x}{v^\beta} \right)^2} .
\end{align}
Setting $x=r$, $y=-r$, we find
\begin{align}
 \langle J^\alpha(-r) J^\beta(r) \rangle
& \sim 
- \frac{g_{\mathrm{e}}}{4\pi^2}
\left( A^{\alpha \beta}_\calB + A^{\beta \alpha}_\calB \right)
\left(\frac{v_\alpha v_\beta}{v^\alpha + v^\beta} \right)^2
\frac{1}{r^2} .
\end{align}
Therefore, we find 
\begin{equation}
 G^{\alpha \beta} = G^{\beta \alpha}
\sim- \frac{e^2}{h}(2 \pi^2)
\left(\frac{v^\alpha + v^\beta}{v^\alpha v^\beta} \right)^2 r^2
 \langle { J^\alpha(-r) J^\beta(r)} \rangle .
 \label{eq:G}
\end{equation}
This is to be compared with Eqs.~(72) and (76) of Ref.~\cite{Rahmani:2012bq} 
or Eq.~(4.4) of Ref.~\cite{Hou:2012hz}
under time-reversal invariance and in the thermodynamic limit:
\begin{equation}
 G^{\alpha \beta} = G^{\beta \alpha} \sim - \frac{e^2}{h}(8 \pi^2)
\frac{1}{v^\alpha v^\beta} r^2
 \langle J^\alpha(-r) J^\beta(r) \rangle .
\end{equation}
That is, our result is different from the result in Ref.~\cite{Rahmani:2012bq} by the factor of
\begin{equation}
\left(\frac{v^\alpha + v^\beta}{2 \sqrt{v^\alpha v^\beta}} \right)^2 .
\end{equation}
Two results are identical if and only if $v^\alpha = v^\beta$.
In the next section, we will demonstrate that the present results~\eqref{eq.JJtau0} and~\eqref{eq:G} are
indeed consistent with numerical simulations.

We shall also emphasize that derivation above only consider the universal and dominant part of the correlation function.
In general there are also nonuniversal parts which decay faster.
That is why in the numerical simulation,
the conductance is extracted from the asymptotic behavior of $G^{\alpha\beta}(r)$ as defined in Eq.~(\ref{eq:G_static}).

\section{Numerical Method}
\label{sec:method}

We note that the linear conductance and the stationary current, 
which are defined by Eq.~(\ref{eq:G_static}) and Eq.~(\ref{eq:J}), respectively, 
are well defined for an infinite system and at infinite time limit.
However, due to the lack of translational invariance, in general it is difficult to simulate a two-wire junction in the thermodynamic limit.
In order to extract the conductance from a finite size calculation, various approaches have been proposed.
One approach is to map an infinite multiwire junction to a finite strip by a conformal mapping \cite{Rahmani:2010jr, Rahmani:2012bq}.
The mapping enables one to relate the linear conductance to the static current-current correlation function in a finite system.
However, in this approach it is necessary to apply an ad hoc mirror boundary condition,
which can be rigorously proved only for the case of non-interacting wires.
Time-dependent DMRG has also been used to simulate the time-dependent local current.
Typically a finite-size system with open boundary condition is studied.
However, it has been show that finite-size effects can be severe \cite{Schneider:2006tn}. 
On one hand, the current will completely reverse its direction after a finite period of time.
On the other hand, the quasistationary current may have oscillation due to the presence of the boundary.
This makes it difficult to obtain a well defined averaged current.
It is known that these finite-size effects can be reduced by using damped boundary condition,
which has been applied to the quantum dot systems connected to metallic leads \cite{daSilva:2008fw} 
and fractional quantum Hall systems \cite{Feiguin:2008jd}.

In this work we develop a framework to evaluate the static correlation functions 
and time-dependent currents for an infinite two-wire system.
Before we outline the major steps, we briefly describe the key points of our approach.
The main strategy of our approach is to work with an infinite system to minimize the finite-size effects but only perform measurements within a finite size window to make the simulation feasible.
To proceed, we always assume that the wave function is in the form of an infinite matrix product state (iMPS).
For translationally invariant systems the corresponding iMPS can be represented by only a few matrices.
In contrast, in principle infinite many different matrices are needed for the iMPS representation 
of the ground state of $H_{\text{junc}}$, since the translation invariance is broken. 
For impurity or interface systems, some of us propose an iMPS ansatz 
that only requires a finite number of matrices in Ref.~\cite{Lo:2019ki}.
The main assumption is that far away from the impurity or interface, 
the wave function should be almost the same as the bulk system without the impurity or interface.
We hence consider a finite size window.
Outside the window, the wave function is assumed to be the same as the corresponding bulk one.
Inside the window, we need to optimize the wave function with an effective Hamiltonian,
which is obtained by attaching infinite boundary conditions (IBCs) at the left and right boundary of the window.
It is shown in Ref.~\cite{Lo:2019ki} that 
the {\em static} correlation functions within the window can be evaluated accurately.
Furthermore, we ensure that the window size is large enough 
such that the current-current correlation has reached its asymptotic behavior.

In this work, we further extend the method to the time-dependent problem by applying the technique proposed in Ref.~\cite{Zauner:2015jl}.
By using this technique, we can evaluate accurately the {\em time-dependent} local currents within a finite-size window.
We show that the amplitude of the residual oscillation in the quasistationary region is extremely small,
and an excellent time-averaged current can be obtained.
In addition, while a small amount of current still reflects at the window boundary, 
the current never totally reverse its direction.
Finally, it will be shown later that a longer quasistationary region can be reached by 
simply enlarging the window size.

Now we are in a position to outline the major steps of our framework.
We start from the iMPS ansatz with window size $2L$ for the two-wire junction:
\begin{eqnarray}
  \label{eq:iMPS_2L}
  & &  |\Psi^{2L}_{\text{junc}}(t) \rangle 
  = \sum_{\vec{n}} \cdots A^{\alpha}_{n_{-L-\frac{3}{2}}}(t) A^{\alpha}_{n_{-L-\frac{1}{2}}} (t)  M_{n_{-L+\frac{1}{2}}}(t)\cdots \nonumber \\
  & & \cdots M_{n_{L-\frac{1}{2}}}(t)
   B^{\beta}_{n_{L+\frac{1}{2}}}(t)  B^{\beta}_{n_{L+\frac{3}{2}}}(t)  \cdots |\vec{n}\rangle, 
\end{eqnarray}
where $A^{\alpha}_i(t)=A^{\alpha}(t)$ and $B^{\beta}_i(t)=B^\beta(t)$ are site-independent $D\times D$ matrices, 
$M^{\alpha}_i(t)$ are site-dependent $D\times D$ matrices, 
$|\vec{n}\rangle = |\dots n_{-\frac{3}{2}} n_{-\frac{1}{2}} n_{\frac{1}{2}} n_{\frac{3}{2}} \cdots \rangle$ is the product basis,
and $D$ is the maximum virtual bond dimension. 
We assume a one site unit cell but it is straightforward to allow a multisite unit cell.
When $t \le 0$,  we assume $|\Psi^{2L}_{\text{junc}}(t \le 0)\rangle$ is the variational ground state of $H_{\text{junc}}$.
When $t \ge 0$, we time evolve $|\Psi^{2L}_{\text{junc}}(0)\rangle$ with $H_{\text{junc}}+H_{\text{bias}}(V)$.

To find $A^{\alpha}(t \le 0)$, $B^\beta(t \le 0)$, and $M_i(t \le 0)$ 
we first use infinite-size DMRG algorithms \cite{McCulloch.2008} to obtain the ground state $|\psi^{\mu}_{\text{wire}} \rangle$ of wire $\mu$ 
with Hamiltonian $H^\mu_{\text{wire}}$ as an iMPS in the mixed canonical form \cite{Schollwock:2011gl}:
\begin{equation}
  |\Psi^{\mu}_{\text{wire}} \rangle   = \sum_{\vec{n}} 
  \cdots \tilde{A}^{\mu} \tilde{A}^{\mu} \lambda \tilde{B}^{\mu} \tilde{B}^{\mu} \cdots |\vec{n}\rangle,
\end{equation}
where $\lambda$ is a $D\times D$ diagonal matrix
while $A^\mu$ and $B^\mu$ are $D\times D$ matrices.
Furthermore, 
$\tilde{A}^\mu$ satisfy the left canonical form constraint
$\sum_{\mu}\tilde{A}^{\mu\dagger}\tilde{A}^{\mu}=I$,
and $\tilde{B}^{\mu}$ satisfy the right canonical form constraint
$\sum_{\mu}\tilde{B}^{\mu}\tilde{B}^{\mu\dagger}=I$ \cite{Schollwock:2011gl}.
We then assume $A^{\alpha}(t\le 0) = \tilde{A}^\alpha$ and $B^\beta(t\le 0)=\tilde{B}^\beta$.
We next construct an effective $2L$-sites Hamiltonian as
\begin{equation}
  H_{\text{junc}}^{2L} = \widetilde{W}_L
  W^{\alpha}_{-L+\frac{1}{2}} \cdots W^{\alpha}_{-\frac{3}{2}} \widetilde{W}_{-\frac{1}{2}} 
  W^{\beta}_{\frac{1}{2}} \cdots  W^{\beta}_{L-\frac{1}{2}} 
  \widetilde{W}_R,
\end{equation}
where $W^{\mu}_i$ is defined in Eq.~(\ref{MPO_bulk}) for the infinite wires,
$\widetilde{W}_{-\frac{1}{2}}$ is defined in Eq.~(\ref{MPO_junc}),
and $\widetilde{W}_L$ ($\widetilde{W}_R$) represents the left (right) infinite boundary condition (IBC).
We note that the left IBC $\widetilde{W}_L$  is constructed from $A^\alpha$ and $W^\alpha$,
while the right IBC $\widetilde{W}_R$ is constructed from $B^\beta$ and $W^\beta$.
The detail of the procedure to construct IBC can be found in Ref.~\cite{Lo:2019ki}.
We use the finite-size DMRG algorithm to obtain the ground state of $H_{\text{junc}}^{2L}$ as a finite MPS:
\begin{equation}
  \sum_{\vec{n}} \widetilde{M}_{n_{-L+\frac{1}{2}}} \cdots \widetilde{M}_{n_{L-\frac{1}{2}}} |\vec{n}\rangle.
\end{equation}
We then assume $M_i(t<0)=\widetilde{M}_i$. 

\begin{figure}[t]
  \includegraphics[width=\columnwidth]{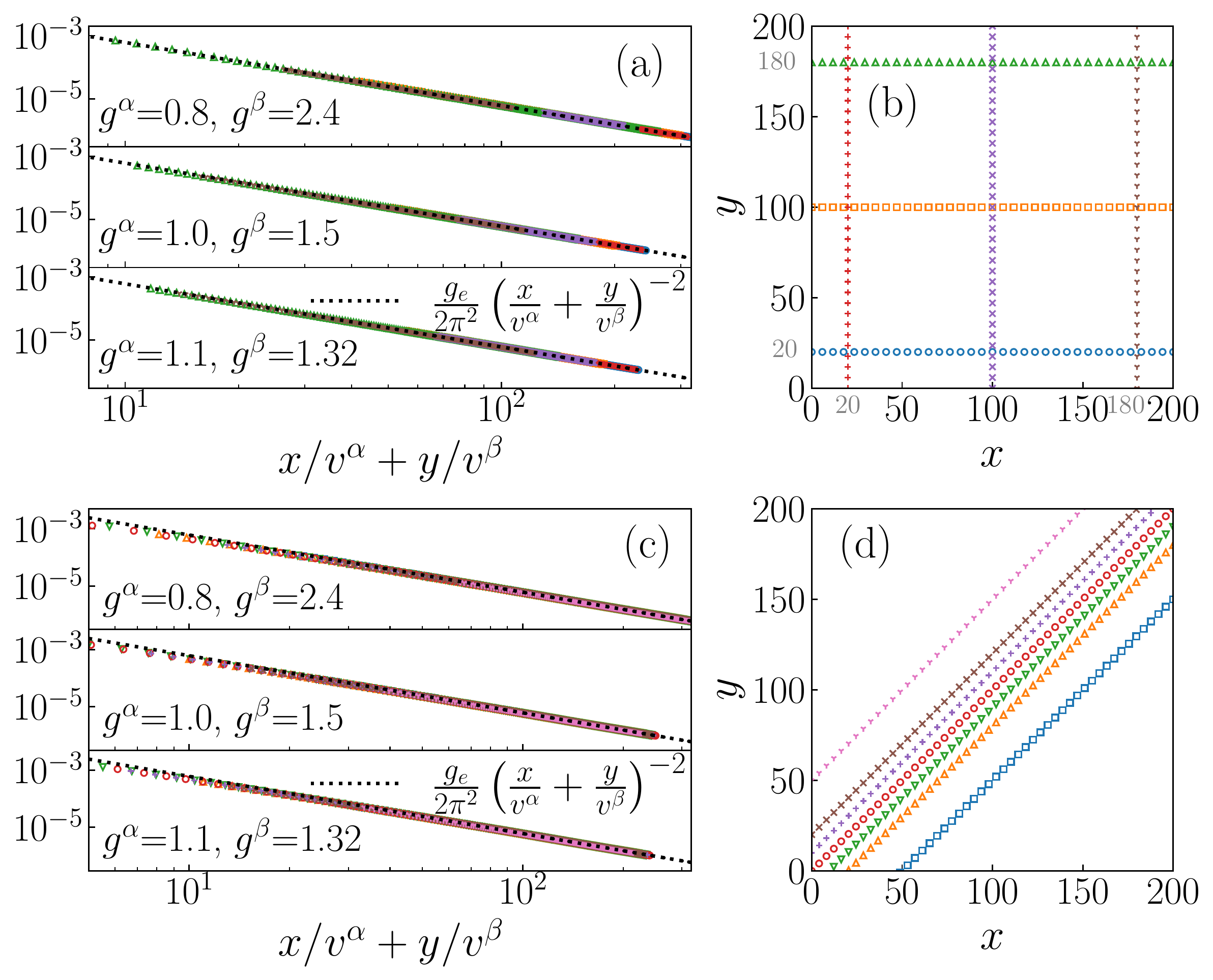}
  \caption{
  Equal-time current-current correlation functions $\langle J^\alpha(y) J^\beta(x) \rangle$ versus $-y/v^\alpha+x/v^\beta$.
  Although three different sets of parameters giving different values of carrier velocities and Luttinger parameters are used,
  they all give the same effective Luttinger parameter $g_{\text{e}}=1.2$.
  The numerical data of $\langle J^\alpha(y) J^\beta(x) \rangle$ are shown in (a) and (c), for the set of coordinates $(x,y)$ as shown
  in (b) and (d), respectively.
  They collapse into a single line in the log-log plot with respect to the scaling variable $-y/v^\alpha+x/v^\beta$, in agreement with
  the theoretical prediction~\eqref{eq.JJtau0} with~\eqref{eq:A_Dirichlet} (dotted line).}
  \label{fig:corr_parallel_ge12}
\end{figure}

At this stage we have obtained $|\Psi^{2L}_{\text{junc}}(t\le 0) \rangle$,
from which we can evaluate the ground-state current-current correlation function $\langle J(-r) J(r) \rangle_{\text{junc}}$.
We note that due to the left and right canonical conditions of the $A$ and $B$ matrices, 
the correlation functions within the window can be evaluated using only $M$ matrices.
We next calculate the position dependent conductance $G^{\alpha \beta}(r)$
and use the asymptotic behavior of $G^{\alpha \beta}(r)$ to estimate the conductance according to Eq.~(\ref{eq:G_static}).
This concludes the time-independent part of the calculation.

To find $A^{\alpha}(t \ge 0)$, $B^\beta(t \ge 0)$, and $M_i(t \ge 0)$, 
we first break the time-evolution operator into products of time-evolution operator with a small time step $dt \ll 1$.
At each time step, we use second-order Suzuki-Trotter decomposition to approximate the evolution operator as products of local gates.
Due to the lack of translational invariance we perform the time evolution in three substeps:
\begin{itemize}
  \item Perform infinite-size TEBD update for $A^\alpha(t)$ and $B^\beta(t)$ 
  with bulk Hamiltonian $H^\alpha_{\text{wire}}$ and $H^\beta_{\text{wire}}$, respectively.
  We note that they are site independent (up to a unit cell) and standard infinite size TEBD update is used.
  \item Perform finite-size TEBD update for $M_i(t)$ with the effective Hamiltonian $H^{2L}_{\text{junc}}$,  
   except $M_{\mp n_{L\pm \frac{1}{2}}}(t)$ at the boundary of the window.
  \item Perform special update for $M_{\mp n_{L\pm \frac{1}{2}}}(t)$ as described in Ref.~\cite{Zauner:2015jl}.
\end{itemize}
At this stage we have obtained $A^{\alpha}(t + dt)$, $B^\beta(t + dt)$, and $M_i(t + dt)$.
We then evaluate $\langle J(r,t+dt)\rangle$ and iterate the procedure.

Some comments are in order.
While the wave function is always represented by an iMPS, only finite numbers of matrices need to be updated at each time step.
In principle the window can co-move with the wave front. Here we fix the window size for simplicity.
There are two main factors that limit the time scale that one can simulate accurately the time-dependent current.
First, TEBD update will eventually break down. Typically, by increasing the maximum virtual bond dimension, a larger time scale can be reached.
Second, due to the finite virtual bond dimension, the reflection at the boundary cannot be complete removed.
One can reduce the amount of reflection by increasing the maximum virtual bond dimension of the IBC or one can increase the window size to delay the time of reflection. 
In this work we always check how our results depend on maximum virtual bond dimension and window size to ensure the convergence of the results.

\section{Time-independent results}
\label{sec:static}

\begin{figure}[t]
  \includegraphics[width=0.9\columnwidth]{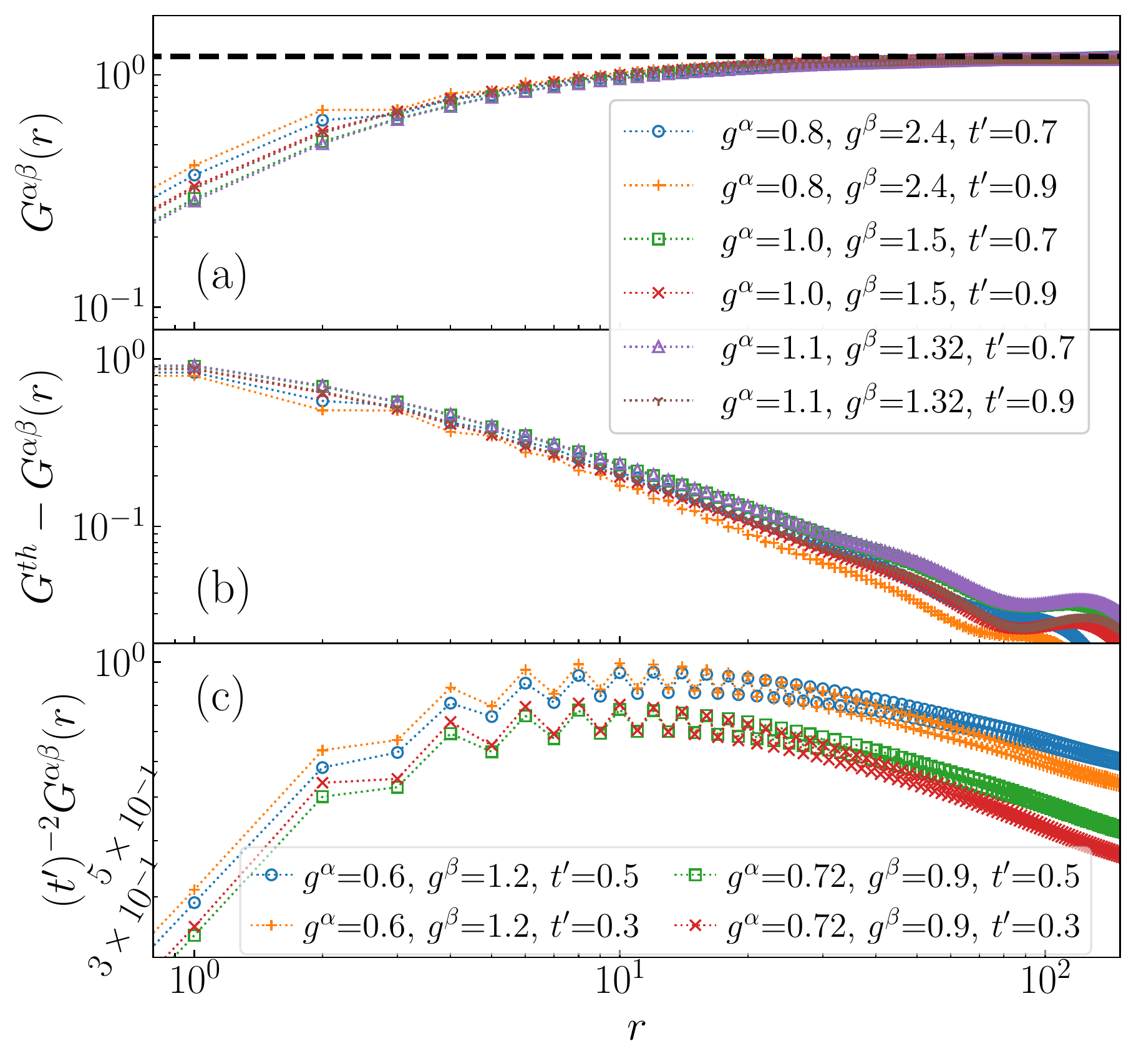}
  \caption{\label{fig:Gab_ge12}
  (a) $G^{\alpha\beta}(r)/\frac{e^2}{h}$ and (b) $[G^{\text{th}}-G^{\alpha\beta}(r)]/\frac{e^2}{h}$ vs $r$ for the case of $g_{\text{e}}=1.2$.
  Black dashed line represents the theoretical prediction.
  (c) $(t^\prime)^{-2}G^{\alpha\beta}(r)/\frac{e^2}{h}$ vs $r$ for the case of $g_{\text{e}}=0.8$.}
\end{figure}

In this section we present our results of the static current-current correlation function and the linear conductance.
We first focus on the case of a two-wire junction with an effective Luttinger parameter $g_{\text{e}} >1$.
In this case it is expected that the system renormalizes to a single fully connected wire with a linear conductance $G^{\text{th}}=g_{\text{e}} e^2/h$,
regardless of the Luttinger parameter of each individual wire and the link strength $t^\prime$.
Specifically we consider three combinations of $g^\alpha$ and $g^\beta$ that give rise to $g_{\text{e}}=1.2$:
(i) $g^\alpha=0.8$ and $g^\beta=2.4$, (ii) $g^\alpha=1.0$ and $g^\beta=1.5$, and (iii) $g^\alpha=1.10,$ and $g^\beta=1.32$.

In the left column of Fig.~\ref{fig:corr_parallel_ge12} we plot the static current-current correlation function $\langle J^\alpha(y)J^\beta(x) \rangle$ 
as a function of $-y/v^\alpha + x/v^\beta$ on a log-log scale, where the variable $x$ and $y$ are restricted to the corresponding lines in the right column.
This choice of the scaling variable is motivated by the theoretical analysis in Sec.~\ref{sec:bosonization}.
Here we set link strength $t^\prime=0.7$, maximum virtual bond dimension $D=800$, and window size $2L=400$.
We observe that the correlation function quickly converges to Eq.~\eqref{eq.JJtau0} (dotted black line) with the condition~\eqref{eq:A_Dirichlet},
regardless of the parameters used, in excellent agreement with the theoretical prediction.
Furthermore, we also perform simulation with $t^\prime=0.9$ and we find that the results are indistinguishable from the results shown above.
We note in passing that this also verifies that the window size and maximum virtual bond dimension are large enough to capture the asymptotic behavior.

\begin{figure}[t]
  \includegraphics[width=\columnwidth]{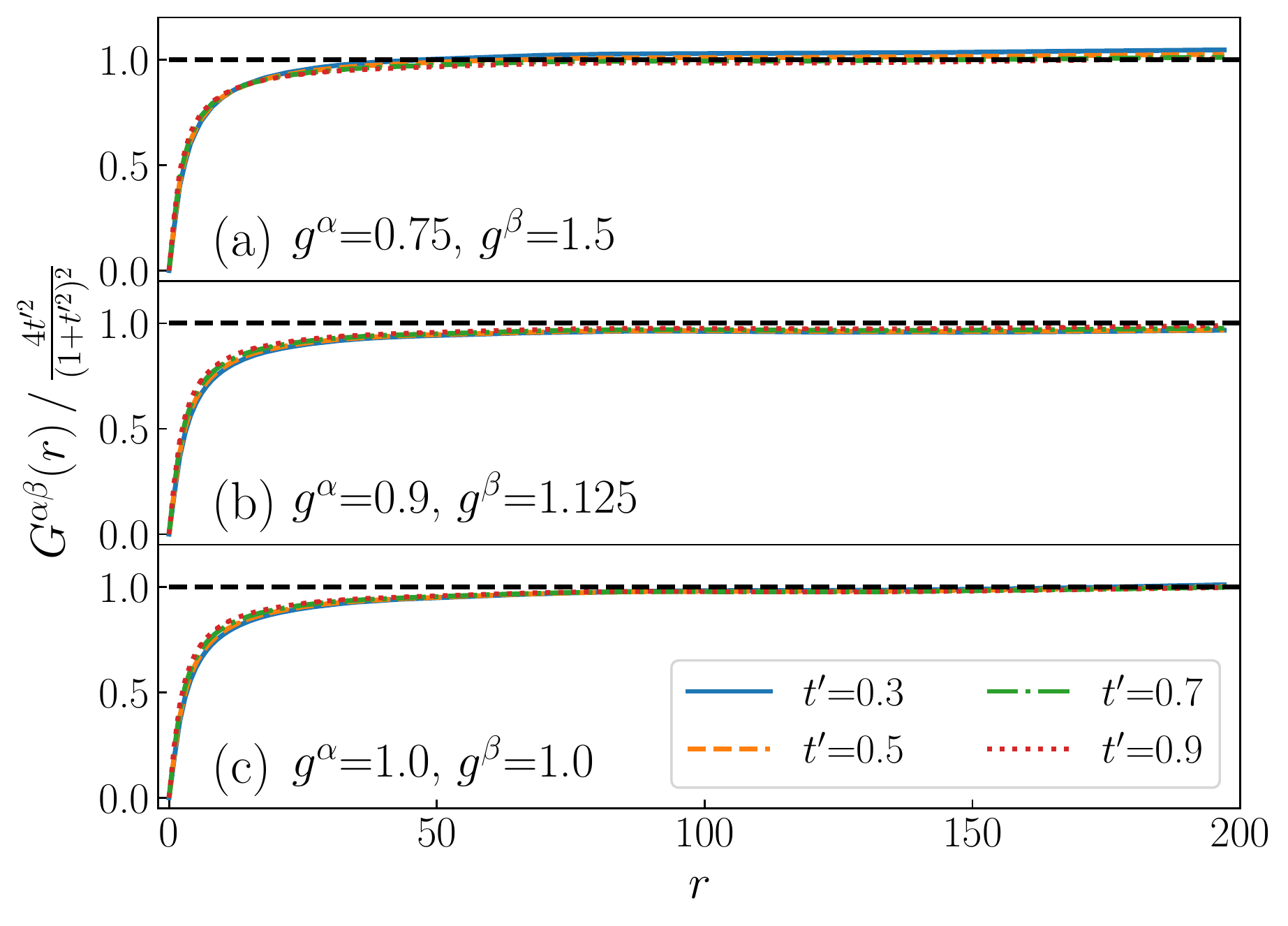}
  \caption{The $r$-dependent dimensionless conductance $G^{\alpha\beta}(r)/\frac{e^2}{h}$ rescaled by $4t^{\prime 2}/(1+t^{\prime 2})^2$ vs $r$ for the cases of $g_{\text{e}}=1$.
  The black dashed lines represent the theoretical prediction.
  \label{fig:evol_tll_Gt} }
\end{figure}

Let us make a more quantitative comparison between the theory and the numerical data, in terms of the conductance
estimated using Eq.~(\ref{eq:G}).
In Fig.~\ref{fig:Gab_ge12}(a) we show the distance dependent conductance $G(r)$ (in units of $e^2/h$) as a function of $r$.
We observe that for all combinations of $g^{\alpha, \beta}$ and $t^\prime$, $G(r)$ quickly approaches the theoretical prediction (dashed black line).
To investigate how $G(r)$ approaches its asymptotic value,
in Fig.~\ref{fig:Gab_ge12}(b) we show $G^{\text{th}} - G(r)$ vs $r$ on a log-log scale.
We find that asymptotically it decays nearly as a power law.
We note that the bump at large distance is due to the finite window effect, which can be eliminated by enlarging the window size.

We next study the case of a two-wire junction with an effective Luttinger parameter $g_{\text{e}} <1$.
In this case it is expected that the system renormalizes to two disconnected wires fixed point with $G^{\text{th}}=0$.
Specifically, we use two combinations of $g^\alpha$ and $g^\beta$: 
(i) $g^\alpha=0.6$, $g^\beta=1.2$ and (ii) $g^\alpha=0.72$, $g^\beta=0.9$ that give rise to $g_{\text{e}}=0.8$.
To probe the disconnected wires behavior we use two values of link strength $t^\prime=0.3$ and $t^\prime = 0.5$.
Smaller $t^\prime$ is used here to ensure that the correlation function can reach its asymptotic behavior within the window.
In  Fig.~\ref{fig:Gab_ge12}(c) we show $ t^{\prime -2} G(r) $ as a function of $r$ on a log-log scale.
We observe a nonuniversal behavior and the value of $G(r)$ depends on $g^{\alpha,\beta}$ and $t^\prime$.
However, asymptotically $G(r)$ always decays as a power law.
While our data is limited by the window size, we expect that $\lim_{r\rightarrow \infty} G(r)=0$,
consistent with the broken wire interpretation.

Finally we investigate the case of $g_{\text{e}}=1$. 
 Here we use three combinations of $g^{\alpha, \beta}$ that give rise to $g_{\text{e}}=1$:
 (a) $g^\alpha=0.75$, $g^\beta=1.5$, (b) $g^\alpha=0.9$, $g^\beta=1.125$, and (c) $g^\alpha=g^\beta=1$.
In Fig.~\ref{fig:evol_tll_Gt} we plot $G^{\alpha \beta}(r)/ G^{\text{th}}$ as a function of $r$.
The ratio approaches one regardless of the parameters used, but we also find a tiny deviation when the wires are interacting.

\section{Time-dependent results}
\label{sec:time}

\begin{figure}[t]
  \includegraphics[width=\columnwidth]{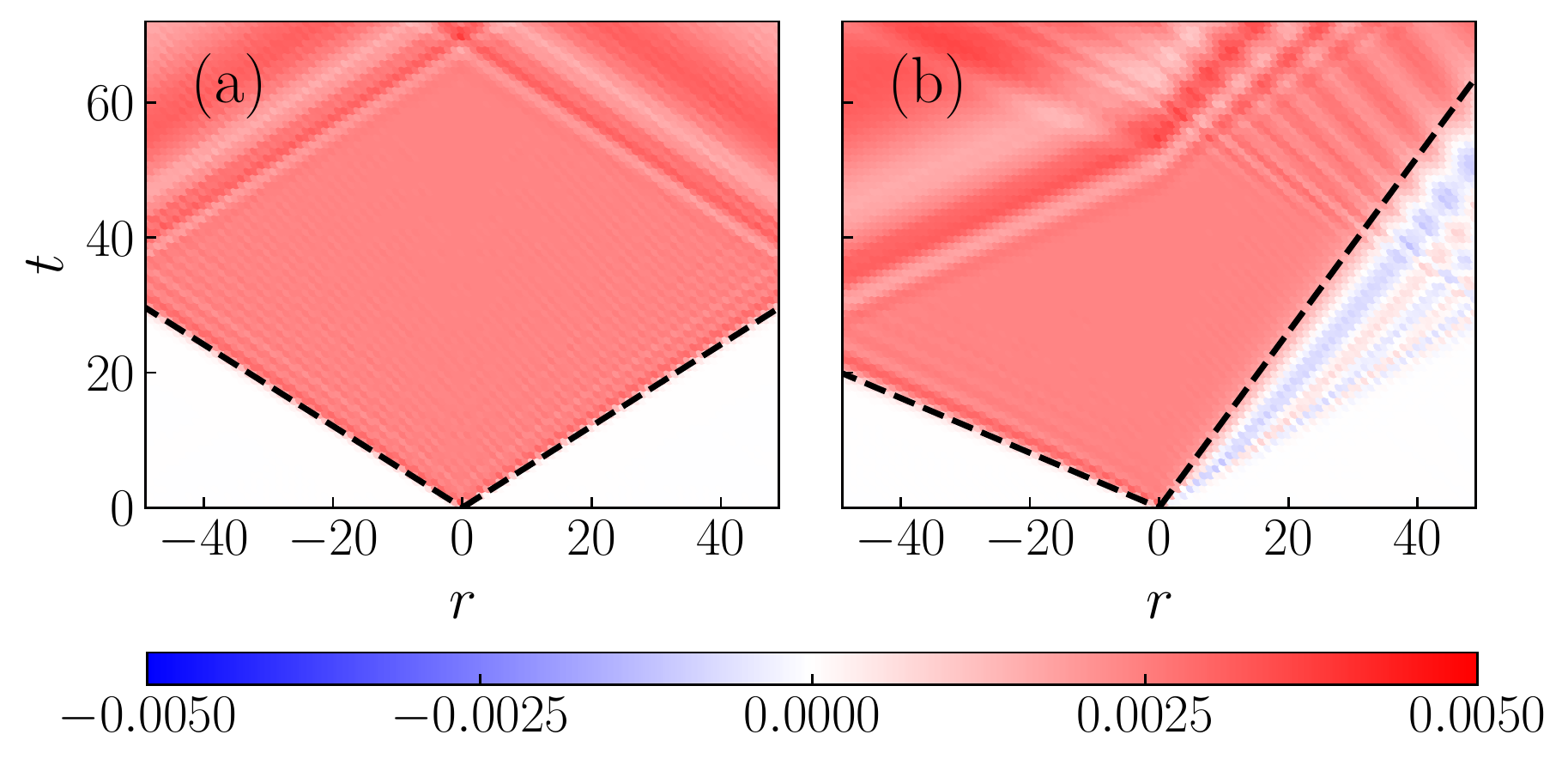} 
  \caption{\label{fig:evol_tll_ge12_0}
  Color density plot of the $J(r,t)/\frac{e}{h}$ in a space-time representation for the case of  (a) $g^\alpha=g^\beta=1.2$ and (b) $g^\alpha=0.8$, $g^\beta=2.4$.
  The dashed lines correspond to $r=\pm v^\mu t$, where $v^\mu$ is the carrier velocity.}
\end{figure}

In this section we present our time-dependent results.
We first benchmark our method with a junction of two equivalent wires with $g^\mu=1.2$. 
The hopping strength $t^\prime$ between wires is set to $0.8$.
We use second order Suzuki-Trotter expansion with $dt=0.002$ to perform the time evolution.
The window size $2L$ is set to 100 and the maximum virtual bond dimension $D$ is 200.
In Fig.~\ref{fig:evol_tll_ge12_0}(a), we plot $\langle J(r,t)\rangle_V$ (in units of $e^2/h$) on the $r-t$ plane after a small bias $V=0.002$ is applied to the system at $t=0$.
It is expected that the current will first appear in a location where the voltage is reversed.
Indeed we observe that the current emerges at the junction ( $r=0$ ) once we turn on the bias.
After that, the fronts propagate in both directions and a light cone is formed.
In Fig.~\ref{fig:evol_tll_ge12_0}(a) we also plot $r=\pm v^\mu t$ as dashed lines, where $v^\mu$ is the carrier velocity.
It is evident that the slope of the light cone agrees well with the carrier velocity.
While the window technique can minimize the reflection at the window boundary,
partial reflection still occurs due to the finite virtual bond dimension.
We find that the front hits the window around $t \approx L/v^\mu \approx 30$. 
However, the partial reflection does not appear until around $t \approx 37$.

\begin{figure}[t]
  \includegraphics[width=\columnwidth]{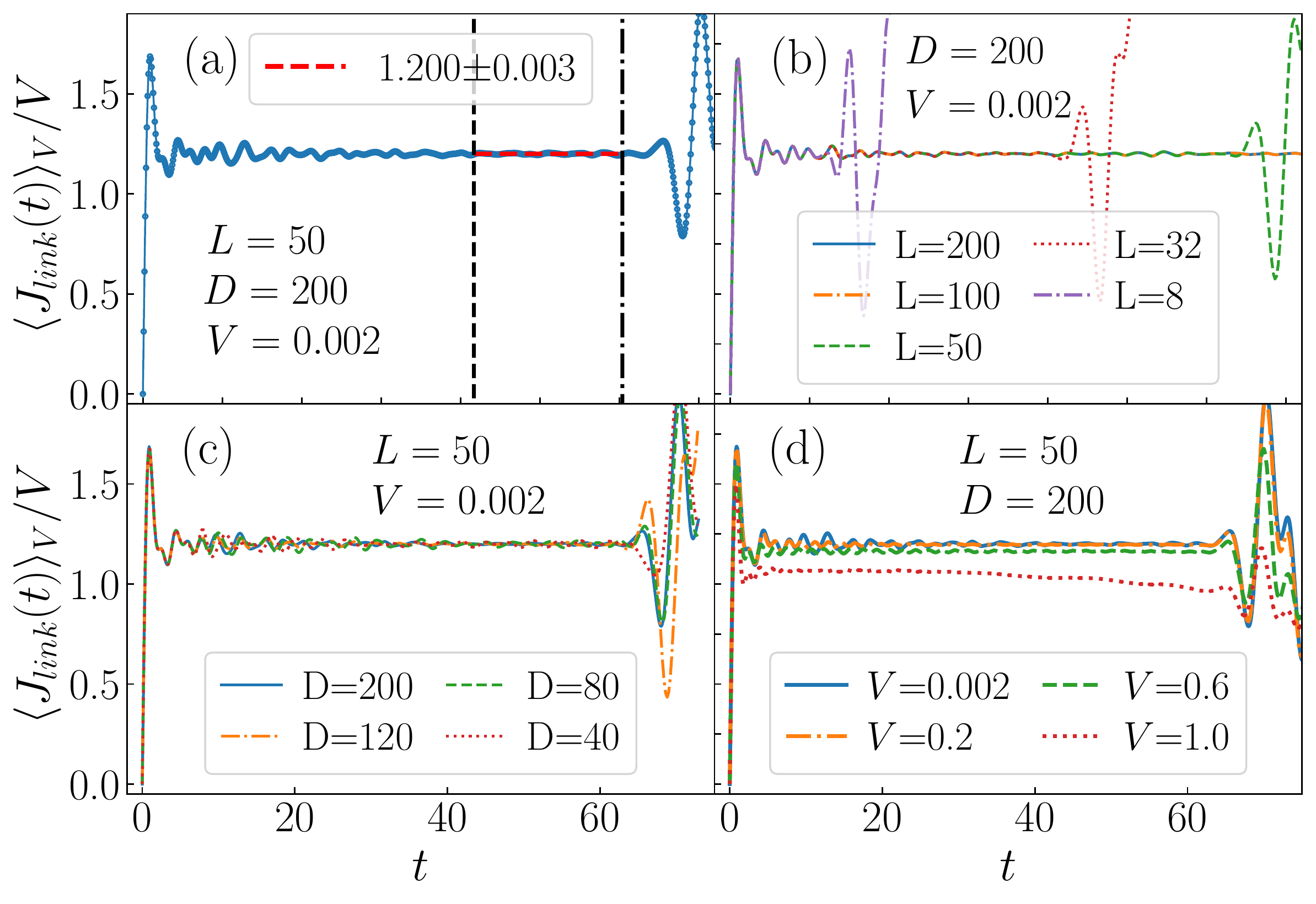}
  \caption{\label{fig:evol_tll_ge12_0_t}
  $\langle J_{\text{link}}(t)\rangle_V/V\frac{e^2}{h}$ with $V=0.002$ vs $t$ for the case of $g^\alpha=g^\beta=1.2$.
  The vertical dashed and dash-dot lines corresponds to $t_1$ and $t_2$, respectively,
  where $[t_1, t_2]$ is the time interval over which the time average of the current is taken.}
\end{figure}
\begin{table}[h!]
  \caption{The numerical results of the conductances and the estimate error in units of $e^2/h$, which are obtained from the time-dependent calculations,
  for the case of $g^\alpha=g^\beta=1.2$, with $V=0.002$.}
  \begin{center}
    \begin{tabular*}{\hsize}{@{}@{\extracolsep{\fill}}cccccc@{}}
    \hline\hline
    \qquad \qquad D & $\overline{J}_{\text{link}}/V$   & $\epsilon$   \\ 
    \hline
  \qquad \qquad 40 \ \   & \ \ $1.199$ \ \ & \ \ $0.009$ \ \  \\
  \qquad \qquad 80 \ \   & \ \ $1.200$ \ \ & \ \ $0.009$ \ \  \\
  \qquad \qquad 120 \ \  & \ \ $1.200$ \ \ & \ \ $0.004$ \ \  \\
  \qquad \qquad 200 \ \  & \ \ $1.200$ \ \ & \ \ $0.004$ \ \  \\
  \hline \hline
  \end{tabular*}
  \end{center} \label{tab:condt}
\end{table}

We next turn our attention to the time-dependent current across the link.
In Fig.~\ref{fig:evol_tll_ge12_0_t}(a) we plot $\langle J_{\text{link}}(t)\rangle_V/V$ (in units of $e^2/h$) as a function of time.
We observe a fast increase from zero once we turn on the bias, followed by the transiently decaying oscillations.
This oscillation is bias dependent and due to the backward scattering and Andreev-type reflection at the junction
\cite{Dolcini:2003, Dolcini:2005, Recher:2006, Pugnetti:2009}.
After that, the current becomes quasistationary until $t \approx 70$.
At this time the fronts of the partially reflected current reach $r=0$ and unphysical oscillations start to emerge.
However, the current across the link never reverses its direction in our simulation. 
This is in contrast to the simulations with open boundary condition,
in which the current will change direction periodically.
It is worth mentioning that the amplitude of residual oscillations in the quasistationary region is extremely small. 
This is because we work with an infinite system and the finite-size induced oscillation is eliminated.
To obtain the averaged current we identify a time interval $(t_1, t_2)$ as follows:
We first set $t_2$ to be a time that is slightly before the reflected front reaches the center.
We then move $t_1$ away from $t_2$ until the estimated error is minimized.
In Fig.~\ref{fig:evol_tll_ge12_0_t}(a) we draw $t_1$ and $t_2$ as vertical dashed and dashed-dotted lines, respectively.
From the time-averaged current we find $\bar{J}_{\text{link}}/V\frac{e^2}{h}=1.200 \pm 0.004$, 
which agrees excellently with the expected results of $G^{\text{th}}=\frac{e^2}{h}g_{\text{e}}(=1.2\frac{e^2}{h})$.

In general, there are two simulation parameters which determine the accuracy of the time-averaged current $\bar{J}_{\text{link}}$.
The window size $L$ determines the maximal time scale before the unphysical oscillations appears.
The maximum virtual bond dimension $D$ determines the amount of reflection as well as the quality of time evolution.
To investigate how our numerical results depend on $L$ and $D$,
we first fix $D$ and run the simulations with various $L$ as shown in Fig.~\ref{fig:evol_tll_ge12_0_t}(b).
We observe that results from different $L$ almost collapse into a single curve before the unphysical oscillation occurs.
Next we fix $L=50$ and run the simulations with various $D$. 
As shown in Fig.~\ref{fig:evol_tll_ge12_0_t}(c), we find that 
both the amplitude of the residual oscillations in the quasistationary region and the amplitude of the unphysical oscillation decrease as $D$ increases.
This is consistent with the picture that quality of time-evolution increases as $D$ increases.
In Table~\ref{tab:condt}, we list our numerical results of $\bar{J}_{\text{link}}/V\frac{e^2}{h}$ and the estimated error, which are obtained with various $D$.
We find that results obtained with $D=40$ already agree well with the theoretical prediction and the estimated error is already small.
Furthermore, the error continues to decrease as $D$ increases. 
It is worth mentioning that this is one of the advantages of the method proposed in this work.
For a similar simulation in Ref.~\cite{AlHassanieh:2006fd},  the tDMRG results are not convergent until $D \ge 300$.


\begin{figure}[t]
  \includegraphics[width=0.9\columnwidth]{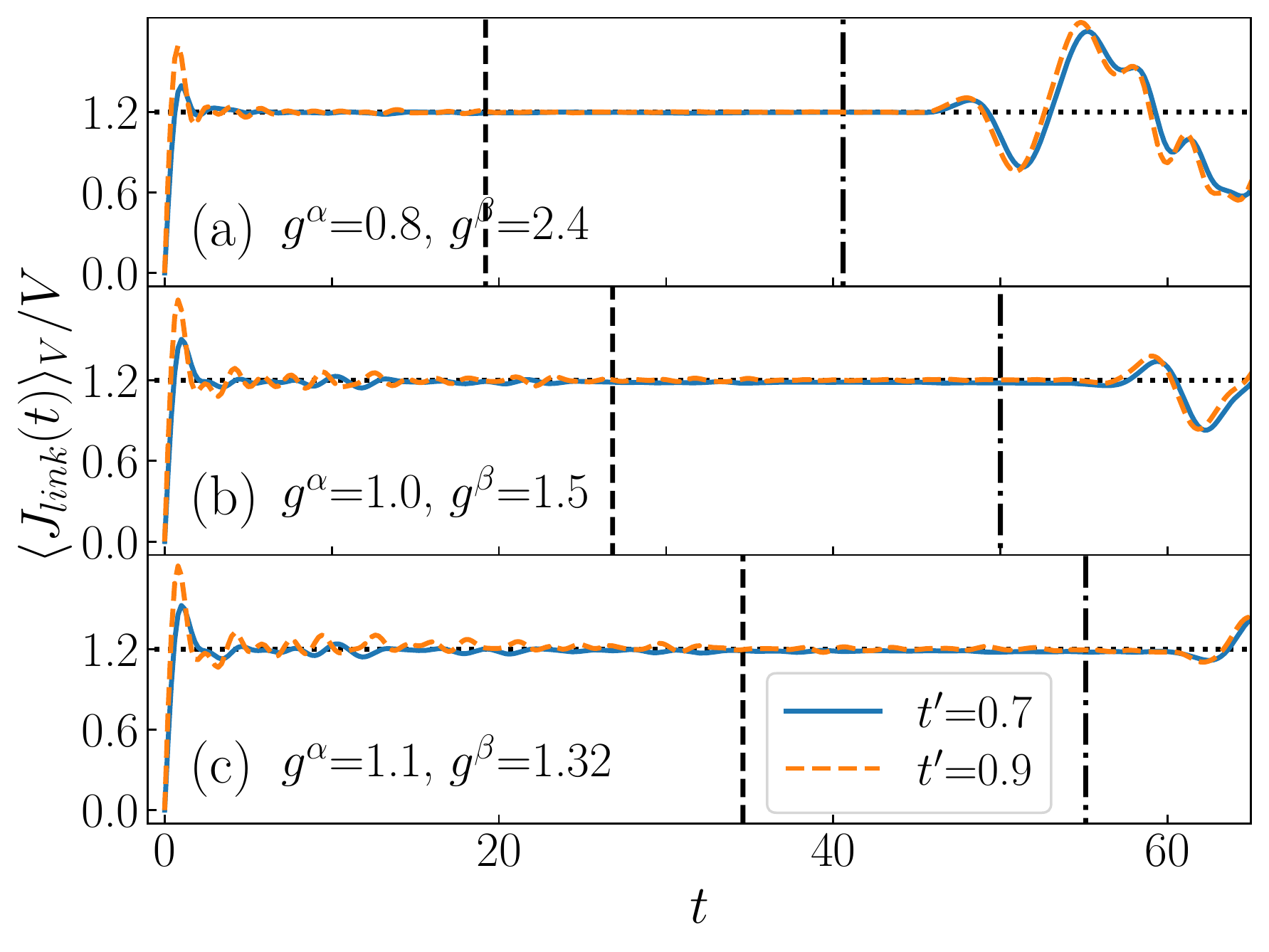}
  \caption{
  	$\langle J_{\text{link}}(t)\rangle_V/V\frac{e^2}{h}$ vs time for the cases of $g^\alpha \neq g^\beta, g_{\text{e}}=1.2$.
  The dotted horizontal lines represent the theoretical predictions.
  The vertical dashed and dash-dot lines correspond to $t_1$ and $t_2$, respectively,
  where $[t_1, t_2]$ is the time interval over which the time average of the current is taken.
  \label{fig:evol_tll_ge12_2}}
\end{figure}
\begin{table}[t]
  \caption{
  The numerical results of the conductances and the estimate error in units of $e^2/h$, which are obtained from the time-dependent calculations,
  for the case of $g^\alpha \neq g^\beta, g_{\text{e}}=1.2$, with $V=0.002$.}
  \begin{center}
  \begin{tabular*}{\hsize}{@{}@{\extracolsep{\fill}}c|ccccc@{}}
    \hline\hline
    & $ G^{\alpha \beta}(t^\prime=0.7)$   &  $G^{\alpha \beta}(t^\prime=0.9)$   \\
    \hline
    \ $g^\alpha=0.8$, $g^\beta=2.4$  \ \,  & \ \ $1.1955\pm0.0008$ \ \  &  \ \ $1.1975\pm0.0013$ \ \   \\
    \ $g^\alpha=1.0$, $g^\beta=1.5$  \ \,  & \ \ $1.1845\pm0.0020$ \ \  &  \ \ $1.2028\pm0.0034$ \ \   \\
    \ $g^\alpha=1.1$, $g^\beta=1.32$ \  & \ \ $1.1783\pm0.0017$ \ \  &  \ \ $1.1998\pm0.0076$ \ \   \\
    \hline \hline
  \end{tabular*}
  \end{center} \label{tab:condt-2g}
\end{table}

We note that while the finite-size induced oscillations are eliminated in the quasistationary regime,
$\langle J_{\text{link}}(t)\rangle_V/V\frac{e^2}{h}$ still has residual oscillations. 
We find that, however, the amplitude of the residual oscillation is almost independent of the bias.
Consequently, for larger bias the residual oscillations for $\langle J_{\text{link}}(t)\rangle_V/V\frac{e^2}{h}$ are invisible.
In contrast for a very small bias the residual oscillation becomes visible, leading to a larger error in $\bar{J}_{\text{link}}/V\frac{e^2}{h}$.
In order to ensure that the system is in the linear response region with minimized error,
in the following we will use $V = 0.002$ when evaluating the linear conductance.

Now we are in a position to study the case of two inequivalent wires with an effective $g_{\text{e}}>1$.
Three combinations of $g^{\alpha, \beta}$ with $g_{\text{e}}=1.2$ are considered:
(a) $g^\alpha=0.8$, $g^\beta=2.4$, (b) $g^\alpha=1.0$, $g^\beta=1.5$, and (c) $g^\alpha=1.1$, $g^\beta=1.32$,
where we use the convention of $g^\alpha \le g^\beta$. 
Two hopping strengths $t^\prime = 0.7$ and $t^\prime = 0.9$ are used.
In Fig.~\ref{fig:evol_tll_ge12_0}(b) we show the time-dependent local current on the $r-t$ plane.
Due to the different propagation velocities, the light cone is asymmetric.
Since a smaller Luttinger parameter implies larger carrier velocity, 
the current reflected by the left boundary will come back first.
In Fig.~\ref{fig:evol_tll_ge12_2}, we show $\langle J_{\text{link}}(t)\rangle_V/V\frac{e^2}{h}$ as a function of time.
We observe that in all cases it converges to the expected result of $G^{\text{th}}=\frac{e^2}{h}g_{\text{e}}(=1.2\frac{e^2}{h})$,
which is denoted by the dotted horizontal line.
By averaging the current between the dashed and the dash dot vertical lines,
we obtain the conductance as summarized in Table \ref{tab:condt-2g}.
The results are highly consistent with the theoretical prediction as well as
the results from the current-current correlation function as presented in Sec.~\ref{sec:static}.

\begin{figure}
  \includegraphics[width=0.9\columnwidth]{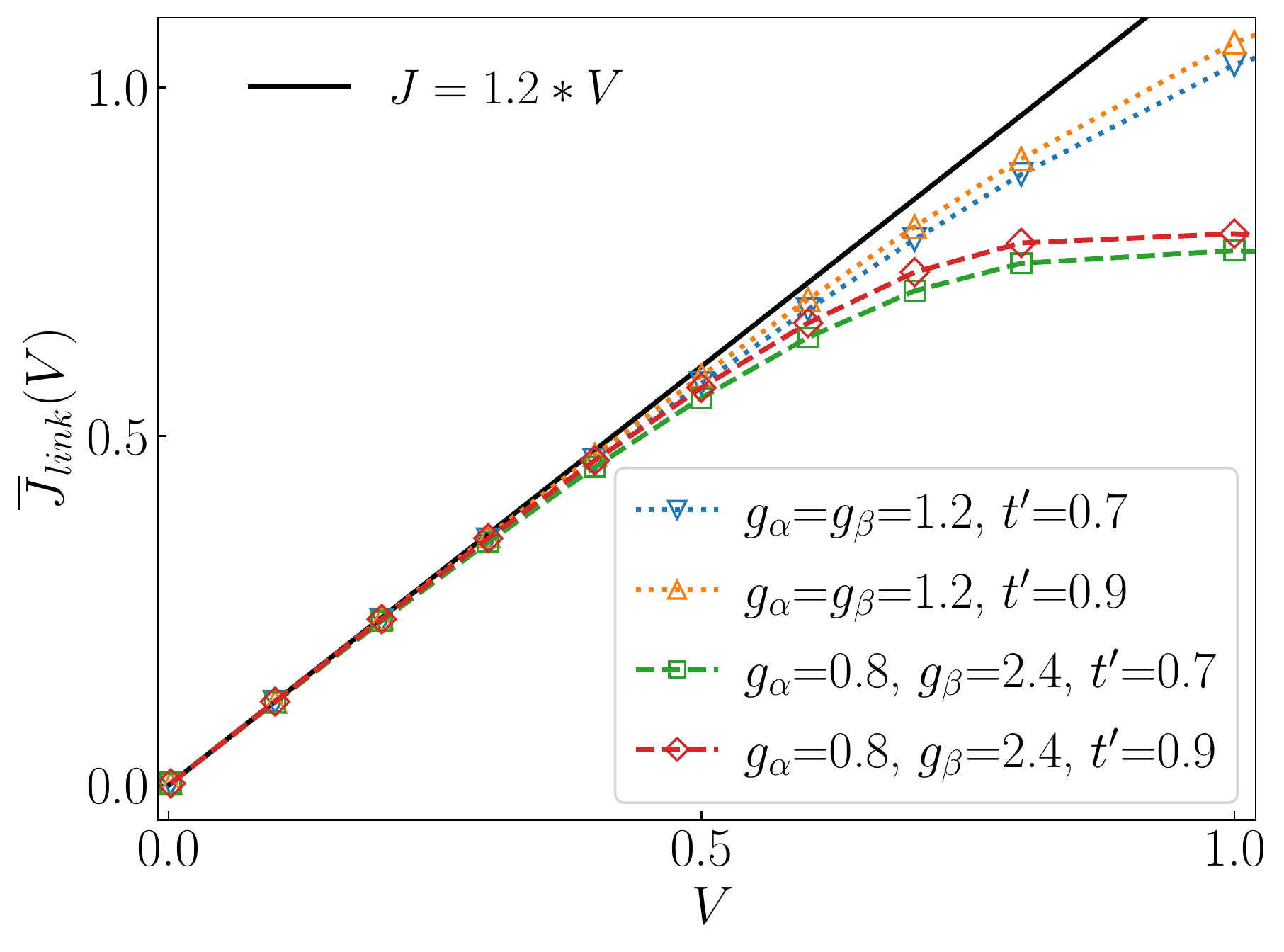} 
  \caption{\label{fig:evol_tll_ge12_GV}
  $\overline{J}_{\text{link}}(V)/\frac{e^2}{h}$ vs bias $V$ for the cases of $g_{\text{e}}=1.2$.
  The solid black line represents the linear response predicted by the theory.
  The dashed and dotted lines are guides to the eye.}
\end{figure}

To probe the nonlinear response, we set $D=200$, $L=50$ and run the simulations with various $V$.
We use $g^\alpha=g^\beta=1.2$ and $g^\alpha=0.8, g^\beta=2.4$, both with $g_{\text{e}}=1.2$.
In Fig.~\ref{fig:evol_tll_ge12_0_t}(d) we show $\langle J_{\text{link}}(t)\rangle_V/V\frac{e^2}{h}$ as a function of time.
A long period of quasistationary region is observed for all cases except when $V$ is close to 1.
This is due to the finite-window effect and can be eliminated by enlarging the window.
In Fig.~\ref{fig:evol_tll_ge12_GV} we show the time-averaged current $\overline{J}_{\text{link}}$ in units of $e^2/h$ as a function of $V$ on a log-log scale.
Two hopping strengths $t^\prime = 0.7$ and $t^\prime = 0.9$ are considered.
We also plot a straight line that corresponds to the linear response $\bar{J}=G^{\text{th}}V=\frac{e^2}{h}g_{\text{e}} V$.
We observe a universal linear response up to $V\approx 0.2$ while at large bias the deviation becomes substantial.
As has been pointed out in Ref.~\cite{Eckel:2010kw}, the deviation at large bias is due to the finite band width of the model.

\begin{figure}
  \includegraphics[width=0.9\columnwidth]{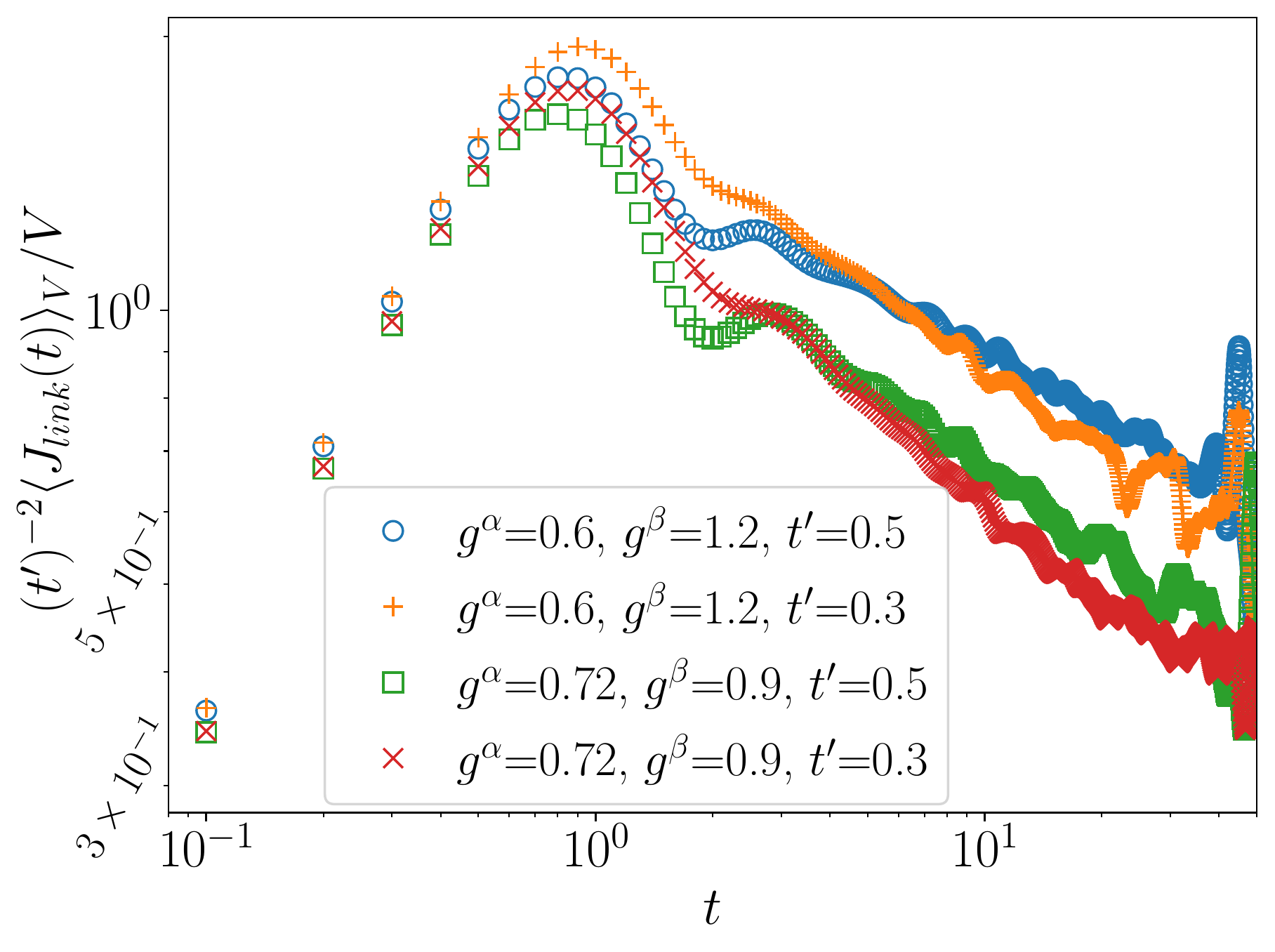}
  \caption{\label{fig:evol_tll_ge08} 
  $(t^\prime)^{-2} \langle J_{\text{link}}(t)\rangle_V/V\frac{e^2}{h}$ vs time on the log-log scale for the cases of $g_{\text{e}}=0.8$.}
\end{figure}

We now investigate the case of $g_{\text{e}}<1$. We consider two combinations
(a) $g^\alpha=0.6$, $g^\beta=1.2$ and (b) $g^\alpha=0.72$, $g^\beta=0.9$, which give rise to $g_{\text{e}}=0.8$.
It is expected that in this case the the junction will flow to the broken wire fixed point with zero conductance.
In order to observe the broken wire behavior within the time window, we use smaller link strength, $t^\prime=0.3$ and $0.5$.
From the perturbation theory we expect that the current might scale as $(t^\prime)^{-2}$.
In Fig.~\ref{fig:evol_tll_ge08} we plot $(t^\prime)^{-2} \langle J_{\text{link}}(t)\rangle_V/V\frac{e^2}{h}$ on a log-log plot.
At short time, we observe a universal rise of the rescaled current.
After that, it decays as a power law but the decay exponent is parameter dependent.
These results indicate that at large time one has $G^{\alpha\beta}=0$, 
which is in agreement with the theoretical predictions and the results of the time-independent method.

\begin{figure}
  \includegraphics[width=0.9\columnwidth]{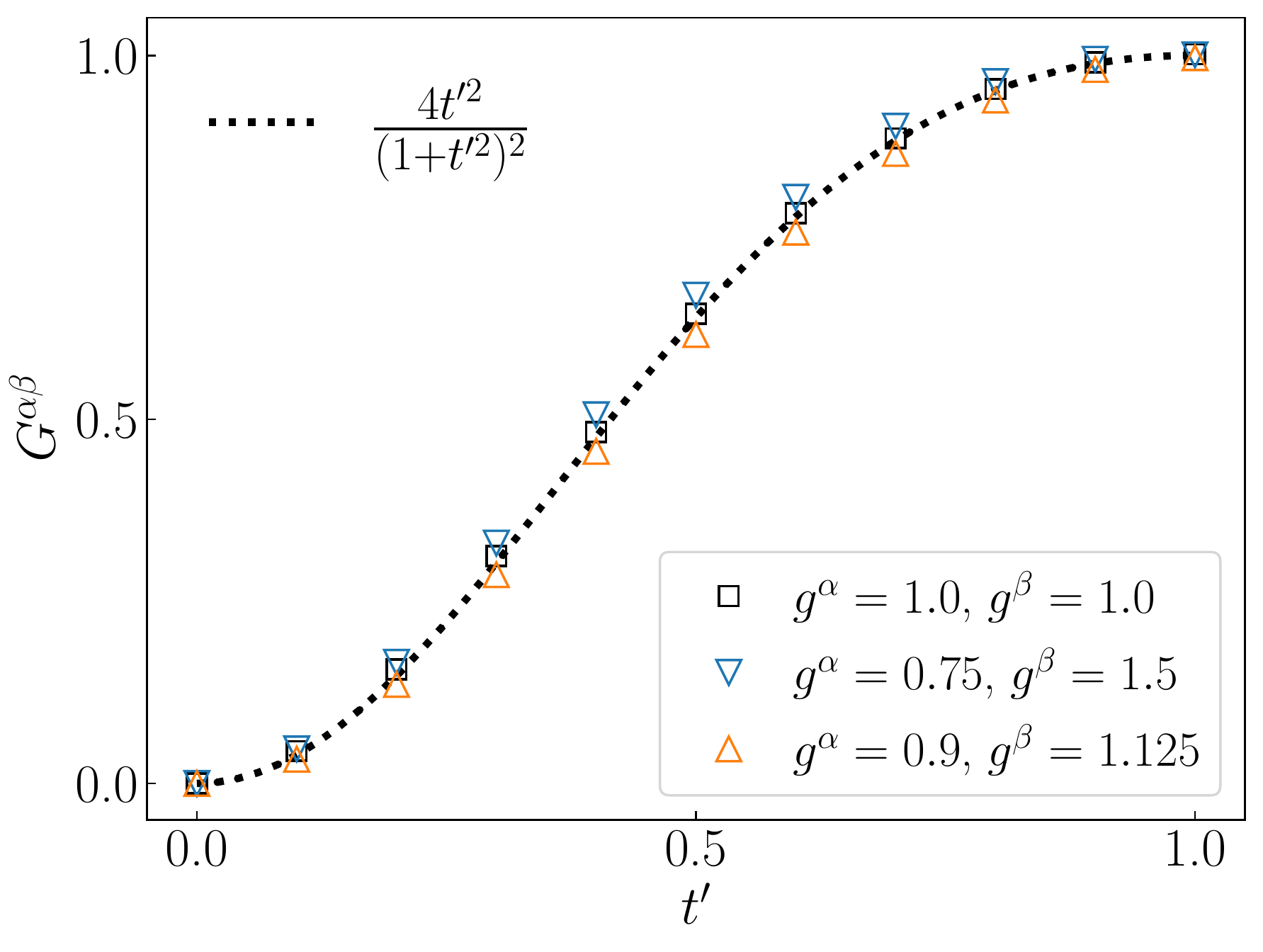}
  \caption{\label{fig:G_ge1} The conductance in units of $e^2/h$ obtained by the time-dependent calculation vs link strength $t^\prime$ for the case of $g_{\text{e}}=1$.
  The dotted line corresponds to the exact solution for the case of noninteracting wires.}
\end{figure}

Finally we investigate the case of $g_{\text{e}}=1$.
In Fig.~\ref{fig:G_ge1} we show our results of the conductance obtained by the time-dependent calculation as a function of link strength.
For junction of equivalent noninteracting wires ($g^\mu=1$), 
the results agree excellently with the theoretical prediction which is shown as the dotted line.
For junction of inequivalent wires ($g^\alpha \neq g^\beta$), the results are very close to the case of noninteracting wires.
However, we do observe a small but systematic deviation.
The results agree with our theoretical conjecture that only a weak change of the conductance is expected in this case.

\section{Summary and discussion}
\label{sec:sum}

In summary, we have developed a novel numerical framework to study the transport properties of two-wire junctions with inequivalent TLL wires,
based on a finite window embedded in an infinite wire.
Within this framework, we implemented two schemes.
In the first scheme,  the linear conductance is extracted from the asymptotic value of the static current-current correlation function,
through a relation which is derived via the bosonization method.
It is worth mentioning that the relation derived in this work corrected a subtle error in Ref.~\cite{Rahmani:2012bq}.
Our result agrees with Ref.~\cite{Rahmani:2012bq} if and only two TLL wires have the same charge velocity, but it is generically different for two inequivalent wires.
In the second scheme, we evaluate the time-dependent local current across the junction after a bias is applied.
The current-voltage characteristic is then extracted by averaging the local current across the junction in the quasistationary region.
In particular, the linear conductance is estimated by applying a small bias.

The main advantage of our schemes is to always work with an infinite system to eliminate the finite-size effects but only perform measurements within a finite window to make the calculation tractable.
We benchmark our schemes against known theoretical results.
For the time-independent calculations, we show that the asymptotic behavior of the current-current correlation function 
can be obtained with a moderate window size. 
Furthermore, the linear conductance extracted agrees excellently with the theoretical prediction.
For the time-dependent calculations, we show that with moderate window size and maximum virtual bond dimension,
a long quasistationary region can be reached. By averaging the current within the quasistationary region
one can accurately determine the current-voltage characteristics.
Furthermore, we obtain the linear conductance by applying a small bias and
the results agree excellently with the theoretical prediction as well as the results via the time-independent method.
It is worth pointing out that computationally it is less demanding to use the time-dependent method to reach the same accuracy.
This is because for the time-independent method, 
it is essential to have high precision results of the correlation function at large distance to obtain its asymptotic behavior.
However, the values of large distance correlations are quite small, making it more demanding to calculate accurately. 
On the other hand, for the time-dependent method accurate results can be obtained as long as
a long quasistationary region with small residual oscillations can be reached.

Some comments are now in order. 
First, it is straightforward to generalize both schemes to study the transport properties of complex multiwire junctions.
For complex geometry, one can study the Y junction with three TLL wires.
By changing the enclosing magnetic flux and the Luttinger parameters, many conductance fixed points can be reached.
It is also possible to include spin degrees of freedom and study the conductance of nanostructures.
Going beyond linear response, it is interesting to study the single impurity Anderson model, where Kondo physics is important.
In summary, our schemes allow one to determine accurately the correlation functions to a very large distance,
simulate the dynamics to a very large time scale, and the results are free of finite size effects.
We believe that these schemes can become important tools to study the transport properties of strongly interacting nanoscopic systems.

\acknowledgments
We acknowledge the support by Ministry of Science and Technology (MOST) of Taiwan through Grant No. 107-2112-M-007-018-MY3 and No. 108-2811-M-007-532. We also acknowledge the support of MEXT/JSPS KAKENHI Grant Nos. JP17H06462 and JP19H01808. Y.T.K. is also supported in part by Department of Education of Guangdong Province (No. 2021KQNCX105) and Key Research Platform and Program in Universities of Department of Education of Guangdong Province (No. 2020GCZX003). The numerical calculation was done using the Uni10 tensor network library (https://uni10.gitlab.io/).

\null\vskip-8mm

\bibliographystyle{apsrev4-2}
\bibliography{ref_new}


\end{document}